# A representation of context-free grammars with the help of finite digraphs

**Krasimir Yordzhev**

Faculty of Mathematics and Natural Sciences, South-West University, Blagoevgrad, Bulgaria

**Email address:**
yordzhev@swu.bg

**Abstract:** For any context-free grammar, we build a transition diagram, that is, a finite directed graph with labeled arcs, which describes the work of the grammar. This approach is new, and it is different from previously known graph models. We define the concept of proper walk in this transition diagram and we prove that a word belongs to a given context-free language if and only if this word can be obtained with the help of a proper walk.

## 1. Introduction

Context-free grammars are widely used to describe the syntax of programming languages and natural languages [1,2]. On the other hand, graph theory is a good apparatus for the modeling of computing devices and computational processes. So a lot of graph algorithms have been developed [4,5]. The purpose of this article is to show how for an arbitrary context-free grammar, we can construct a finite digraph, which describes the work of the grammar. As is well known, any finite state machine (deterministic or nondeterministic finite automaton) and any pushdown automaton can be described by a transition diagram, which is essentially a finite directed graph with labeled arcs [2]. The finite digraph, which we will construct in this work and who will simulate a given context-free grammar will be different from previously known graph models [1,2].

## 2. The Main Definitions and Notations

If
$$\Sigma = \{x_1, x_2, \ldots, x_p\}$$

is a finite set, then with $\Sigma^*$ we will denote the free monoid over $\Sigma$, ie the set of all sequences $x_{i_1} x_{i_2} \ldots x_{i_n}$, including the empty one which we denote with $\varepsilon$. Elements of $\Sigma^*$ is called *words*, and any subset of $\Sigma^*$ (*formal*) *language*.

*Formal* (or *generative*) *grammar* $\Gamma$ is a triple

$$\Gamma = (N, \Sigma, \Pi),$$

where $N$ and $\Sigma$ are finite sets, $N \cap \Sigma = \emptyset$, which are called, respectively, *alphabet of nonterminals* (or *variables*) and *alphabet of terminals*, where $\Pi$ is a finite subset of the Cartesian product $(N \cup \Sigma)^* \times (N \cup \Sigma)^*$, and we will write $u \to v$ if $(u, v) \in \Pi$ (see [3]). We will call $\Pi$ the set of *productions* or *rules*.

Let $\Gamma = (N, \Sigma, \Pi)$ be a formal grammar and let $w, y \in (N \cup \Sigma)^*$. We will write $w \to y$, if there are $z_1, z_2, u, v \in (N \cup \Sigma)^*$, such that $w = z_1 u z_2$, $y = z_1 v z_2$ and $u \to v$ is a production of $\Pi$. We will write $w \Rightarrow y$, if $w = y$, or there are $w_0, w_1, \ldots, w_r$ such that $w_0 = w$ and $w_i \to w_{i+1}$ for all $i = 0, 1, \ldots, r-1$. The sequence $w_0 \to w_1 \to \cdots \to w_r$ is called a *derivation* of length $r$.

For every $A \in N$, a subset $L(\Gamma, A) \subseteq \Sigma^*$ defined by

$$L(\Gamma, A) = \{\alpha \in \Sigma^* \mid A \Rightarrow \alpha\}$$

is called the *language* generated by the grammar $\Gamma$ with the *start symbol* $A$ (see [3]).

The formal grammar $\Gamma = \{N, \Sigma, \Pi\}$ is called *context-free* if all the productions of $\Pi$ are of the form $A \to \omega$, where $A \in N$, $\omega \in (N \cup \Sigma)^*$. The language $L$ is called *context-free* if there is a context-free grammar $\Gamma = (N, \Sigma, \Pi)$ and a nonterminal $A \in N$, such that $L = L(\Gamma, A)$.

*Transition diagram* is called a finite directed graph, all of whose arcs are *labelled* by an element of a semigroup (see also [2]).

If $\pi$ is a walk in a transition diagram, then the *label* of this walk $l(\pi)$ is the product of the labels of the arcs that make up this walk, taken in passing the arcs.

Let $\Gamma = (N, \Sigma, \Pi)$ be a context-free grammar and let

$$N' = \{A' \mid A \in N\} \quad \Sigma' = \{x' \mid x \in \Sigma\}.$$

We consider the monoid $T$ with the set of generators $N \cup \Sigma \cup N' \cup \Sigma'$, and with the defining relations

$$A'A = \varepsilon, \quad x'x = \varepsilon, \qquad (1)$$

where $A' \in N'$, $A \in N$, $x' \in \Sigma'$, $x \in \Sigma$ and $\varepsilon$ is the empty word (the identity of the monoid $T$). We set by definition $(x')' = x$, $(A')' = A$, if $\omega = z_1 z_2 \ldots z_k \in T$, then $\omega'$ will mean $z_k' z_{k-1}' \ldots z_1'$ and $\varepsilon' = \varepsilon$.

Let

$$R = \Sigma^* \times T = \{(\omega, z) \mid \omega \in \Sigma^*, z \in T\}.$$

In $R$ we introduce the operation "∘" as follows: If $(\omega_1, z_1), (\omega_2, z_2) \in R$, then

$$(\omega_1, z_1) \circ (\omega_2, z_2) = (\omega_1 \omega_2, z_2 z_1). \qquad (2)$$

It is easy to see that $R$ with this operation is a monoid with identity $(\varepsilon, \varepsilon)$.

## 3. Finite Digraphs and Context-free Grammars

**Definition 1.** For the context-free grammar $\Gamma$, we construct a transition diagram $H(\Gamma)$ with the set of vertices $V = \{u_A, v_A \mid A \in N\}$ and the set of arcs $E \subseteq V \times V$ which is made as follows:

1. For each production $A \to \alpha$, where $A \in N$, $\alpha \in \Sigma^*$ there is an arc from $u_A$ to $v_A$ with a bel$(\alpha, \varepsilon)$ (see Fig. 1);

2. For each production $A \to \alpha Bz$, where $A, B \in N$, $\alpha \in \Sigma^*$, $z \in (N \cup \Sigma)^*$ there is an arc from $u_A$ to $u_B$ with a label$(\alpha, z)$;

3. For each production $A \to z_1 B_1 \alpha B_2 z_2$ where $A, B_1, B_2 \in N$, $\alpha \in \Sigma^*$, $z_1, z_2 \in (N \cup \Sigma)^*$, there is an arc from $v_{B_1}$ to $u_{B_2}$ with a label$(\alpha, B_2' \alpha')$;

4. For each production $A \to zB\alpha$, where $A, B \in N$, $\alpha \in \Sigma^*$, $z \in (N \cup \Sigma)^*$ there is an arc from $v_B$ to $v_A$ with a label$(\alpha, \alpha')$;

5. There are no other arcs in $H(\Gamma)$ except described in $1 \div 4$.

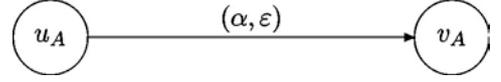

**Figure 1.** Production $A \to \alpha$, $A \in N$, $\alpha \in \Sigma^*$ and the corresponding part of the transition diagram $H(\Gamma)$.

If the production is of the form $A \to \alpha_0 B_1 \alpha_1 B_2 \alpha_2 \ldots \alpha_{k-1} B_k \alpha_k$, where $\alpha_i \in \Sigma^*$ ($i = 0, 1, \ldots k$), $A, B_j \in N$ ($j = 1, 2, \ldots, k$), then the corresponding part of the transition diagram is shown in Figure 2

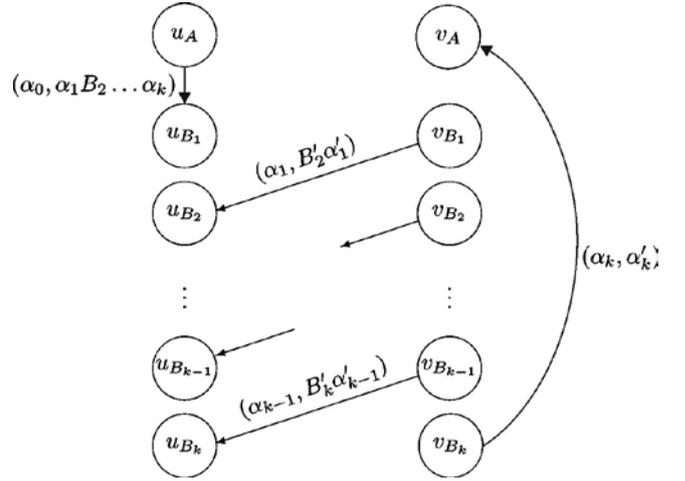

**Figure 2.** Production $A \to \alpha_0 B_1 \alpha_1 B_2 \alpha_2 \ldots \alpha_{k-1} B_k \alpha_k$ and the corresponding part of the transition diagram $H(\Gamma)$.

**Definition 2.** A walk $\pi \in H(\Gamma)$ will be called a *proper walk* if:

1. For each $A \in N$ the arcs from $u_A$ to $v_A$ (if there exist) are proper walks;

2. Let $\pi_1, \pi_2, \ldots, \pi_k$ be proper walks, $A \in N$ and let:
- there is an arc $\rho$ from the vertex $u_A$ to the start vertex of $\pi_1$;
- for every $i = 1, 2, \ldots, k-1$ there are arcs $\sigma_i$ from the final vertex of $\pi_i$ to the start vertex of $\pi_{i+1}$;
- there is an arc $\tau$ from the final vertex of $\pi_k$ to the vertex $v_A$.

If $l(\rho \pi_1 \sigma_1 \pi_2 \sigma_2 \ldots \sigma_{k-1} \pi_k \tau) = (\alpha, \varepsilon)$ for some $\alpha \in \Sigma^*$, then the walk

$$\pi = \rho \pi_1 \sigma_1 \pi_2 \sigma_2 \ldots \sigma_{k-1} \pi_k \tau$$

is the proper walk (see Figure 3).

3. There are no other proper walks in $H(\Gamma)$ except described in 1 and 2.

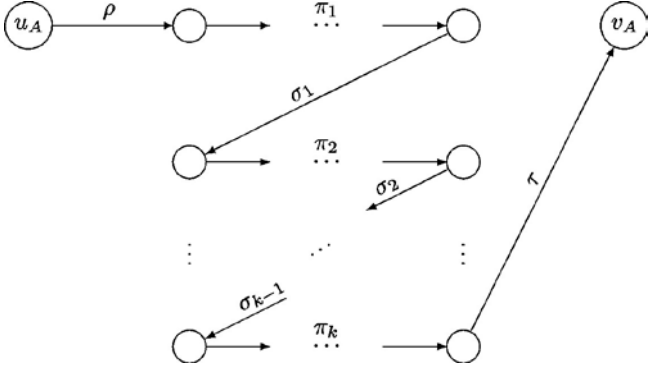

*Figure 3.* *A proper walk in $H(\Gamma)$.*

The following theorem is true.

**Theorem 1** Let $\Gamma = (N, \Sigma, \Pi)$ be a context-free grammar and let $A \in N$. Then $\alpha \in L(\Gamma, A)$ if and only if there is a proper walk in $H(\Gamma)$ with the start vertex $u_A$, the final vertex $v_A$ and with a label $(\alpha, \varepsilon)$.

Proof. <u>Necessity</u>. Let $\alpha \in L(\Gamma, A)$. Then there is a derivation $A \Rightarrow \alpha$. The proof we will make by induction on the length of the derivation.

Let a production $A \to \alpha$ exists. Then according to Definition 1, item 1 there exists an arc $u_A v_A$ with a label $(\alpha, \varepsilon)$.

Suppose that for all $A \in N$ and for all $\alpha \in L(\Gamma, A)$ for which there is a derivation $A \Rightarrow \alpha$ no longer than $t$, there is a proper walk with the start vertex $u_A$, the final vertex $v_A$ and labeled $(\alpha, \varepsilon)$.

Let $\alpha \in L(\Gamma, A)$, and there is a derivation $A \Rightarrow \alpha$ of length $t + 1$. Let $A \to \alpha_0 B_1 \alpha_1 \ldots B_k \alpha_k$ be the first production of this derivation, where $\alpha_i \in \Sigma^*$, $B_j \in N$, $i = 0,1,\ldots k$, $j = 1,2,\ldots k$. Then there is a derivation

$$A \to \alpha_0 B_1 \alpha_1 \ldots B_k \alpha_k \Rightarrow \alpha_0 \beta_1 \alpha_1 \ldots \beta_k \alpha_k = \alpha,$$

where $\beta_i \in L(\Gamma, B_i)$ and for all $\beta_i$ there exist derivations $B_i \Rightarrow \beta_i$ with the length less than or equal to $t$, $i = 1,2,\ldots k$. By the inductive hypothesis, for all $i = 1,2,\ldots,k$, there are proper walks $\pi_i$ with the start vertices $u_{B_i}$, the final vertices $v_{B_i}$ and labeled $(\beta_i, \varepsilon)$.

For the first production $A \to \alpha_0 B_1 \alpha_1 \ldots B_k \alpha_k$ of the derivation $A \Rightarrow \alpha$ we have:

a) According to Definition 1, item 2 there exists an arc $\rho$ from $u_A$ to $u_{B_1}$ with a label $(\alpha_0, \alpha_1 B_2 \ldots B_k \alpha_k)$.

b) According to Definition 1, item 3 for all $i = 1,2,\ldots k-1$ there are arcs $\sigma_i$ from $v_{B_i}$ to $u_{B_{i+1}}$ with a label $(\alpha_i, B'_{i+1}\alpha'_i)$.

c) According to 1, item 1 there is an arc $\tau$ from $v_{B_k}$ to $v_A$ with a label $(\alpha_k, \alpha'_k)$.

We consider the walk $\pi = \rho\pi_1\sigma_1\pi_2\ldots\sigma_{k-1}\pi_k\tau \in H(\Gamma)$. For this walk, we have:

$l(\pi) =$

$= (\alpha_0, \alpha_1 B_2\alpha_2 \ldots B_k\alpha_k) \circ (\beta_1, \varepsilon) \circ (\alpha_1, B'_2\alpha'_1)$
$\quad \circ (\beta_2, \varepsilon) \circ \cdots (\alpha_{k-1}, B'_k\alpha'_{k-1})(\alpha_k, \alpha'_k)$

$= (\alpha_0\beta_1\alpha_1 \ldots \beta_k\alpha_k, \alpha'_k B'_k\alpha'_{k-1} \ldots B'_2\alpha'_1\alpha_1 B_2 \ldots \alpha_{k-1} B_k\alpha_k)$

$= (\alpha, \varepsilon)$

Since $\pi_1, \pi_2 \ldots \pi_k$ are proper walks then by definition 2 $\pi$ is a proper walk in $H(\Gamma)$ with the start vertex $u_A$, the final vertex $v_A$ and labelled $l(\pi) = (\alpha, \varepsilon)$. The necessity is proved.

<u>Sufficiency</u>. Let $\pi$ be a proper walk in $H(\Gamma)$ with the start vertex $u_A$, the final vertex $v_A$ and a label $l(\pi) = (\alpha, \varepsilon)$. If the length of $\pi$ is equal to 1, then $\pi$ is an arc and this is possible if and only if there is a production $A \to \alpha$, ie $\alpha \in L(\Gamma, A)$.

Suppose that for all $A \in N$ and for all proper walks with the start vertex $u_A$, the final vertex $v_A$, length less than or equal to $t$ and with a label $(\alpha, \varepsilon)$ is satisfied $\alpha \in L(\Gamma, A)$. Let $\pi$ be a proper walk with the start vertex $u_A$, the final vertex $v_A$, labelled $(\alpha, \varepsilon)$ and length $t + 1 > 1$. Since $\pi$ is a proper walk, then there is a vertex $B_1 \in N$ such that the first arc $\rho$ in the walk $\pi$ starts at $u_A$ and finishes at $u_{B_1}$. Hence there is a production $A \to \alpha_0 B_1\alpha_1 B_2 \ldots B_k\alpha_k$ in $\Gamma$ such that $l(u_A u_{B_1}) = (\alpha_0, \alpha_1 B_2 \ldots B_k\alpha_k)$ for some $B_i \in N$, $\alpha_j \in \Sigma^*$, $i = 2,3,\ldots,k$, $j = 0,1,\ldots,k$. According to Definition 2, $\pi$ is given by $\pi = \rho\pi_1\sigma_1\pi_2\ldots\sigma_{s-1}\pi_s\tau$, where $\pi_i$ are proper walks with the start vertices $u_{C_i}$ and the final vertices $v_{C_i}$ for some $C_i \in N$ (obviously $C_1 = B_1$), $i = 1,2,\ldots,s$, arcs $\sigma_i = v_{C_i} u_{C_{i+1}}$ from $v_{C_i}$ to $u_{C_{i+1}}$, $i = 1,2,\ldots,s-1$, and an arc $\tau = v_{C_s} v_A$ from $v_{C_s}$ to $v_A$. Let $l(\sigma_i) = (\gamma_i, C'_{i+1}\gamma'_i)$, $\gamma_i \in \Sigma^*$, $i = 1,2,\ldots,s-1$, $l(\tau) = (\gamma_s, \gamma'_s)$, $\gamma \in \Sigma^*$. Then we obtain

$l(\pi) =$

$= l(u_A u_{C_1}) \circ l(\pi_1) \circ l(\sigma_1) \circ l(\pi_2) \circ \ldots \circ l(\sigma_{s-1}) \circ l(\pi_s) \circ l(\tau)$

$= (\omega, \gamma_s, C_s, \gamma_{s-1} \ldots C_2, \gamma_1, \alpha_1 B_2 \ldots B_k\alpha_{k'}),$

where $\omega = \alpha$ and $\gamma'_s C'_s \gamma'_{s-1} \ldots C'_2 \gamma'_1 \alpha_1 B_2 \ldots B_k\alpha_k = \varepsilon$. Having in mind (1), it is easy to see that this is possible if and only if $s = k$, $C_i = B_i$ for $i = 2,3,\ldots,k$ and

$\gamma_j = \alpha_j$ for $j = 1, 2, \dots, k$.

Therefore $\pi = \rho\pi_1\sigma_1 \dots \sigma_{k-1}\pi_k\tau$, where $\pi_i$ are proper walks with the start vertices $u_{B_i}$ and the the final vertices $v_{B_i}$ $(i = 1, 2, \dots, k)$; $\sigma_i$ are arcs from $v_{B_i}$ to $u_{B_{i+1}}$ $(i = 1, 2, \dots, k-1)$; $\rho$ is an arc from $u_A$ to $u_{B_1}$; $\tau$ is an arc from $v_{B_k}$ to $v_A$ and there are $\beta_i \in \Sigma^*$ $(i = 1, 2, \dots, k)$ such that $l(\pi_i) = (\beta_i, \varepsilon)$. By the inductive hypothesis, we have $\beta_i \in L(\Gamma, B_i)$ ie there exist derivations $B_i \Rightarrow \beta_i$ for all $i = 1, 2, \dots, k$. Hence, we have the following derivation

$A \to \alpha_0 B_1 \alpha_1 B_2 \dots B_k \alpha_k \Rightarrow \alpha_0 \beta_1 \alpha_1 \beta_2 \dots \beta_k \alpha_k = \alpha$,

which implies that $\alpha \in L(\Gamma, A)$. The theorem is proved.

## 4. Context-free Grammars in Chomsky Normal Form

Discussed above model is particularly useful when applied to some context-free grammars having a special form, for example, when the grammar is in Chomsky normal form.

**Definition 3.** Let $\Gamma$ be a context-free grammar without the empty word $\varepsilon$ in which all productions are in one of two simple forms, either:

1. $A \to BC$, where A, B, and C, are each nonterminals, or

2. $A \to x$, where A is a nonterminal and x is a terminal.

Such a grammar is said to be in *Chomsky Normal Form.*

As is well known [2], every context-free language *L* can be generated by a grammar in Chomsky normal form.

An algorithm that recognizes whether a word $\alpha \in \Sigma^*$ belongs to the language *L*, that is generated by a grammar in Chomsky normal form. This algorithm works in time $O(n^2)$, where $n = |\alpha|$. Described in section 3 graph model will help us to understand in detail this algorithm for more accurate description.

If the production is of the form $A \to BC$, i.e. it is one of the productions of grammar in Chomsky normal form, then the appropriate fragment of the transition diagram is shown in Figure 4.

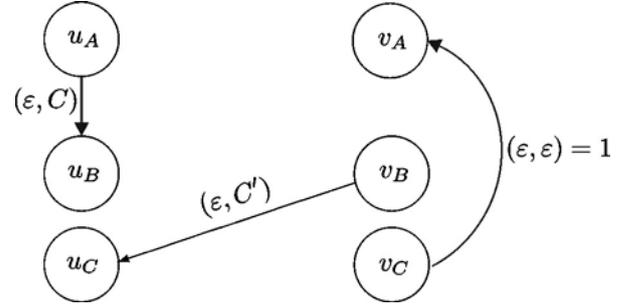

**Figure 4.** *Production $A \to BC$ and the corresponding part of the transition diagram $H(\Gamma)$.*

Obviously, if we consider a context-free grammar in Chomsky normal form, then the corresponding transition diagram is in quite simple and convenient form.

## 5. Conclusions and future work

Described in this article a graph model of an arbitrary context-free grammar is new and original. This is different from the familiar until now graph models.

Some polynomial algorithms for testing the inclusion of any regular or linear language in a group language are described in [7]. These algorithms are based on the search for a cycle in the corresponding transition diagram. This idea can be used for any context-free language L by the transition diagram described in Section 3. The difficulty here is to find, in general, the proper walk, which corresponds to any given word $\alpha \in L$. We hope this open problem to be solved in the near future.

The research is partly supported by the project SRP-B3/13, funded by the Ministry of Education, Youth and Science (Bulgaria) and South-West University "Neofit Rilsky", Blagoevgrad.